# On the superconducting phase fluctuations above $T_c$ in cuprates

A. Lascialfari[1,2], A. Rigamonti[1] and L. Romanò[3]

[1] *Department of Physics "A.Volta", University of Pavia, Via Bassi n.6, I-27100 Pavia, Italy*

[2] *Department of Physics, Università degli studi di Milano, Via Celoria 16, I-20133 Milano, Italy*

[3] *Department of Physics, University of Parma, Parco Area delle Scienze 7A, I-43100 Parma, Italy*

**Abstract**

*In this brief report an attempt is made for a "mise-a-point" of the subject of the phase fluctuations of the superconducting order parameter above $T_c$ in cuprates, particularly as they appear in underdoped compounds. Measurements of torque magnetometry, Nernst effect and isothermal diamagnetic magnetization curves published in the last years are taken into consideration. Although by different experimental approaches and in different magnetic field ranges, it can be stated that vortex-antivortex excitations and phase fluctuations among islands of local non-zero order parameter lacking of long range coherence do occur in a relevant temperature range above $T_c$, particularly in underdoped compounds. The role of the diamagnetic magnetization curves on approaching $T_c$ from above in opening the field with clear signature is remarked, while enlightening comparison with other approaches appear possible.*

The first observation of fluctuating diamagnetism (FD) related to superconducting fluctuations[1] somewhat superimposed to the "conventional" Ginzburg-Landau (GL) contribution, was achieved on a LSCO underdoped single crystal by Lascialfari et al.[2] a long ago. The justification of "unconventional" FD was given by resorting to a theory based on the occurrence of mesoscopic metastable regions above $T_c$, with local non-zero order parameter modulus $|\psi|$ and generation of vortex-antivortex pairs. The strong phase fluctuations were argued to prevent the percolation into a long-range ordered superconducting state. The theoretical picture was based on the extension of an inspiring idea[3] developed in the attempt to justify isothermal magnetization curves above $T_c$ previously reported in underdoped YBCO oriented powders[4]. The indication of extra-diamagnetism above $T_c$ in LSCO was independently supported by scanning superconducting quantum interference microscopy[5]. It is remarked that the presence of mesoscopic "islands" at $|\psi|\neq 0$ above $T_c$ has the consequence of enhanced Nernst effect[6]. It must be added that in a study of precursor diamagnetism in underdoped (aligned powder) LSCO by Carballeira et al.[7], the authors had found discrepancies between "conventional" Prange theory and the experimental data. On the other hand, the low field diamagnetic effect that gives an upturn field $H_{up}$ in fluctuating magnetization $M_{dia}(H, T=\text{const})$ curves[2,4] was not observed in that work. Subsequently, Cabo et al.[8] did observe anomalous FD in underdoped LSCO (and in $Pb_{55}In_{45}$ compounds) and explained the experimental findings by invoking the $T_c^{loc}(\mathbf{r})$ dependence, namely $T_c$ in-homogeneity. $T_c$ in-homogeneity and similar effects due to disorder (for instance substitutionally induced) do induce enhanced diamagnetism above the bulk $T_c$. However, as it was stressed [9], the crucial point allowing to discern among the vortex-antivortex presence and the $T_c$ in-homogeneity contributions to FD (although can simultaneously be both present in general) is the behaviour of the upturn field $H_{up}(T)$ at which $|M_{dia}|$ initiates to decrease on increasing field (above $T_c$). In fact, in case of precursor diamagnetism related to $T_c$ in-homogeneity and/or disorder $H_{up}(T)$ decreases as T is increased since it mimics the critical field $H_{c1}$. On the contrary in case of FD due to vortex-antivortex pairs the theory predicts that $H_{up}(T)$ increases as T is increased. This crucial point will be better illustrated later on.

Now we turn to take into consideration the torque magnetization measurements. In a detailed paper, Li et al.[10] reported a series of torque magnetization (M) measurements on a group of superconducting cuprates, in particular including $La_{2-x}Sr_xCuO_4$ and $Bi_2Sr_{2-x}La_yCuO_6$ (Bi2201). Focusing on the magnetization data above the superconducting transition temperature $T_c$, the experimental results for M vs H, as well as the Nernst signals detected by the same group, are analyzed at the aim to address the following:

a)  a diamagnetic term $M_{dia}$, superimposed to the background paramagnetic contribution, arises at a temperature well above $T_c$;
b)  the temperature range where $M_{dia}$ is observed agrees with the one where vortex Nernst signals are detected as well;
c)  $M_{dia}$ is not linear in the field over a broad temperature range and it is suppressed by intense H;

d)      the diamagnetic response above $T_c$ in cuprates is qualitatively (and quantitatively) different from the one that superconducting fluctuations[1] induce in low-$T_c$ (BCS) superconductors;

e)      the enhanced diamagnetism above $T_c$ in cuprates is qualitatively different (and of different nature) from the precursor diamagnetism characteristic of disordered systems with diffuse transition having "local" $T_c^{loc}(\mathbf{r})$ above the bulk $T_c$;

f)      the strong T and H dependence of the diamagnetic term $M_{dia}$ arises from the pair condensate which above $T_c$ survives as a diluted vortex liquid and with short phase-correlation length.

It is noted that in the paper by Li et al.[10] our magnetization data described above, their analysis and the basic conclusions pointing out the occurrence of phase fluctuations, have been only marginally discussed, essentially limited to the comparison with the effects of diffuse transition in disordered systems.

Now we emphasise how the crucial point allowing to discern among the vortex-antivortex presence and the $T_c$ in-homogeneity contributions to FD *is the behaviour of the upturn field $H_{up}(T)$ at which $|M_{dia}|$ initiates to decrease on increasing field (above $T_c$)*. In fact, in case of precursor diamagnetism related to $T_c$ in-homogeneity and/or disorder, then $H_{up}(T)$ decreases as T is increased since it mimics the critical field $H_{c1}$. On the contrary in case of FD due to vortex-antivortex pairs the theory[2,4] predicts that $H_{up}(T)$ increases as T is increased (see the following).

At the purpose to present conclusive evidence about the occurrence of superconducting phase fluctuations in underdoped cuprates beyond the conventional GL character as evidenced in the diamagnetic magnetization curves above $T_c$, now we just recall the results obtained in Sm-based underdoped cuprates[11]. In Figures 1a and 1b we report the isothermal diamagnetic curves above $T_c$ and the temperature dependence of the upturn field detected in the correspondent conditions.

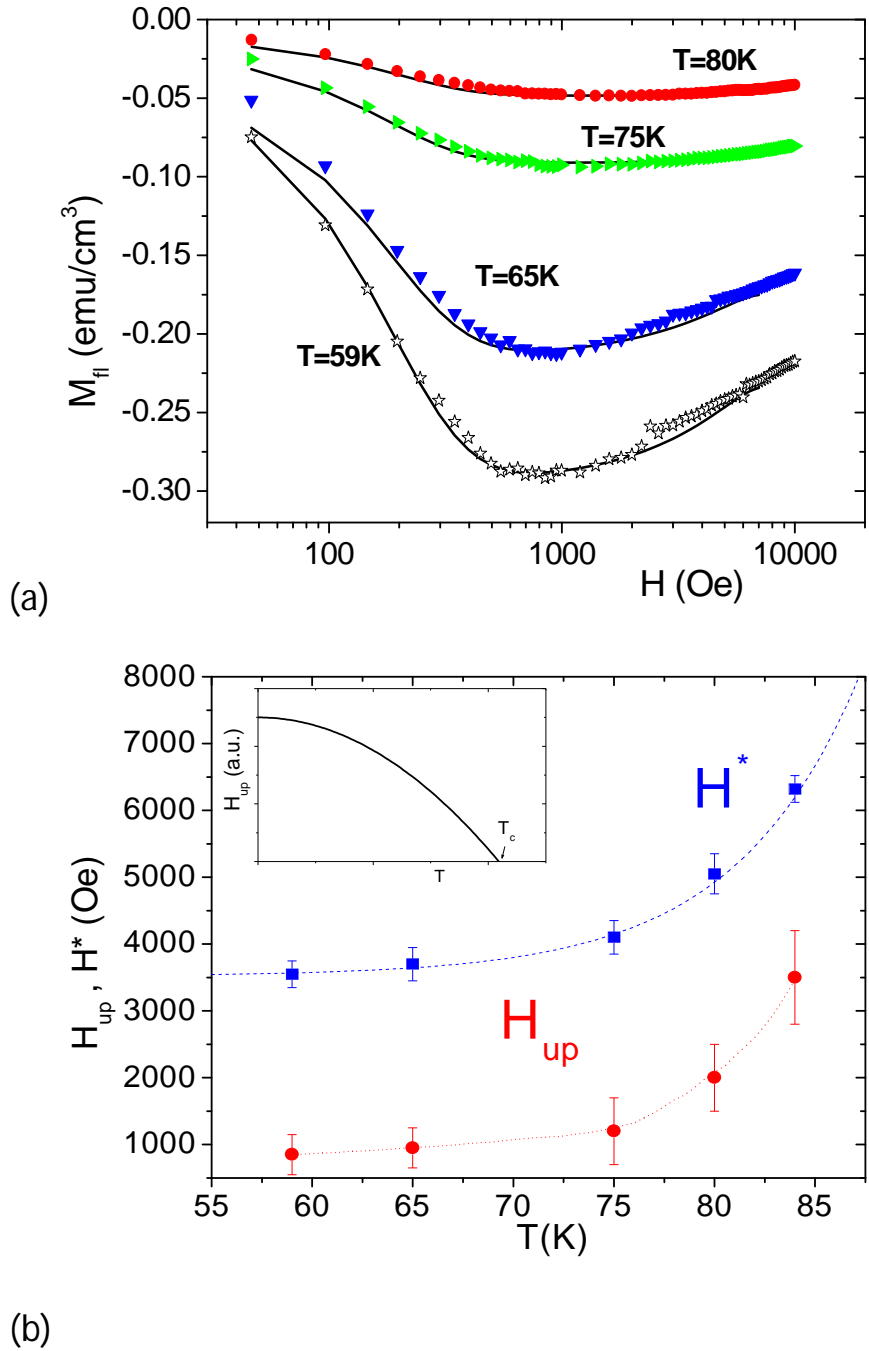

**Fig.1** (a) Representative isothermal magnetization curves $M_{dia}$ vs H in underdoped $SmBa_2Cu_{2.85}Al_{0.15}O_7$ superconductor having $T_c(0) \approx 56.5$ K. The solid lines are the best fits according to the numerical integration of Eq.3 in the text. (b) Upturn field $H_{up}$ and characteristic field $H^*$ (see Eq.3) as a function of temperature in $SmBa_2Cu_{2.85}Al_{0.15}O_7$. The solid line in the inset represents a sketchy behaviour of the upturn field in the case of diffuse transition, namely when $H_{up}$ mimics $H_{c1}$ and therefore it decreases with increasing temperature.

It should be remarked that the value of the upturn field (in the range 500-2000 Oe) has little to do with the field H′≈ $\varepsilon\Phi_0/ 4\xi_0^2$ where the first order fluctuation correction (Gaussian and zero-dimensional approximation) roughly predicts the expected GL quenching of the SF (see Ref.1). In fact, while in BCS superconductors $\varepsilon\Phi_0/4\xi_0^2$ is typically around 100 Oe, in cuprates because of the small coherence length $\xi_0$, very high magnetic field are requested to quench the FD and for $\varepsilon$= (T-$T_c$)/$T_c$ typically in the range $10^{-2}$-$10^{-3}$ one would expect H′≈10 Tesla. At variance, the upturn field in Fig.1b is the direct consequence of FD according to the picture of "islands" at $|\Psi|\neq 0$, with strong phase fluctuations. The upturn field increasing on increasing temperature should be considered the "smoking gun" that the non-conventional GL enhanced diamagnetism is fluctuation-related and does not track a possible site-dependence of the transition temperature.

Finally for a convincing evaluation of our conclusions it is appropriate to summarize the main steps of the theoretical description that allowed us to claim the occurrence of the phase fluctuations in the light of the experimental findings recalled above.

In the GL-Lawrence-Doniach functional

$$F_{LD}[\psi]=\Sigma_l\int d^2r(\alpha|\psi_l|^2+\frac{\beta}{2}|\psi_l|^4+\frac{\hbar^2}{4\pi^2 m}|\left(\nabla_{//}-\frac{2ie}{c\hbar}A_{//}\right)\psi_l|^2+\Im|\psi_{l+1}-\psi_l|^2) \tag{1}$$

we keep $|\Psi|^2$ constant and consider only the dependence on the phase, thus writing

$$F_{LD}[\theta]=\frac{1}{s}\Sigma_l\int d^2r\left\{J_{//}(\nabla_{//}\theta-\frac{2ie}{c\hbar}A_{//})^2+J_\perp[1-\cos(\theta_{l+1}-\theta_l)]\right\}, \tag{2}$$

The second derivative of the free energy with respect to H yields the field-dependent susceptibility

$$\chi=-\frac{k_B T}{s\Phi_0^2}\frac{1}{\langle 1+2n\rangle}\left\{\frac{\left(1+\left(\frac{H}{H^*}\right)^2\delta\right)^2}{n_{vor}}+s^2\gamma^2(1+n)n\left[1+\left(\frac{H}{H^*}\right)^2\delta\right]+0.27L^2\frac{J_{//}}{s}\left(\frac{2\pi}{\Phi_0}\right)^2\left(\frac{H}{H^*}\right)^2\delta\right\} \tag{3}$$

By means of numerical integration the diamagnetic magnetization as a function of the field is obtained in the form of the solid lines in Fig.1a, in correspondence to fit parameters (H* and $\delta$) given in details in Ref.11. Since the explanation of our SQUID magnetization data has been obtained in the framework of the interpretative theory [2,4,1] based on the idea of non-percolating superconducting regions above $T_c$ with strong phase fluctuations, one can conclude that the issues (a)-(f) summarized in the presentation of the torque magnetization data are totally consistent with the scenario we pointed out.

On the other hand, it as to be remarked that our magnetometry data refer to the low magnetic field range, while the data obtained by Li et al.[10] on the basis of torque magnetization measurements refer to strong fields.

## References

[1] A. Larkin and A. A. Varlamov, "*Theory of Fluctuations in Superconductors*" (Oxford University Press, New York, 2005)

[2] A. Lascialfari, A. Rigamonti, L. Romano, A. A. Varlamov, I. Zucca, Phys. Rev. B 68, 100505 (2003)

[3] A. Sewer and H. Beck, Phys. Rev. B 64, 014510 (2001)

[4] P.Carretta, A.Lascialfari, A.Rigamonti, A.Rosso and A.Varlamov, Phys. Rev. B 61, 12420 (2000).

[5] I. Iguchi, T.Yamaguchi and A. Sagimoto, Nature 412, 420 (2001)

[6] Y. Wang, L. Li and N.P. Ong, Phys. Rev. B 73, 024510 (2006)

[7] C. Carballeira, J. Mosqueira, A. Revcolevschi, and F. Vidal, Phys. Rev. Lett. 84, 3157 (2000)

[8] L. Cabo, F. Soto, M. Ruibal, J. Mosqueira, and F. Vidal, Phys. Rev. B 73, 184520 (2006)

[9] A.Rigamonti, A.Lascialfari, L. Romanò, A.Varlamov and I. Zucca, J. Superconductivity 18, 163 (2005), ISSN:0896-1107, 1572.9605 (on-line).